\documentclass[%
reprint,
amsmath,amssymb,
aps,
]{revtex4-2}
\usepackage{float}
\usepackage{epstopdf}
\usepackage{amsmath}
\usepackage{graphicx}
\usepackage{dcolumn}
\usepackage{bm}

\begin{document}
	
	\preprint{APS/123-QED}
	
	\title{Formation of bound states in the continuum in double trapezoidal grating}

	\author{Yuhang Ruan,$^{1}$ Jicheng Wang,$^{1,2}$ YIpan Lou, Zheng-Da Hu,$^{1}$ and Yixiang Wang$^{1}$}
	
	\altaffiliation{wangyixiang@jiangnan.edu.cn
		}
	
	\affiliation{$^{1}$School of Science, Jiangsu Provincial Research Center of Light Industrial Optoelectronic Engineering and Technology, Jiangnan University, Wuxi 214122, China\\
	$^{2}$State Key Laboratory of Millimeter Waves, Southeast University, Nanjing, 210096, China
	}%

	\date{\today}
	
	\begin{abstract}
		In the field of optics, bound state in the continuum (BIC) has been researched in many photonic crystals and periodic structures due to a strong resonance and an ultrahigh \textit{Q} factor. Some designs of narrowband transmission filters, lasers, and sensors were proposed based on excellent optical properties of BIC. In this paper, we consider symmetrical rectangular grating structure firstly, then cut off the corner of one of the gratings, the Fano peak of quasi-BIC can be observed in the spectrum. After that, we further change the tilt parameter of the other grating, which minimizes the Fano line width. In the momentum space, the process of structural change corresponds to topological charges split from $\textit{q}=1$ into two half charges $\textit{q}=1/2$. We analyze guided mode resonance (GMR) excitation of the grating structure, and discuss the dispersion relations in the waveguide layer with the position of BIC in energy bands. In addition, the reflectance spectrum is found to exhibit asymmetric line-shapes with different values of the asymmetry parameters, $M_{1}$ and $M_{2}$. BIC is transformed into quasi-BIC as the symmetry of the structure is broken. Thus, a large Goos–Hänchen shift can be achieved as a result of asymmetrical Fano line shape in quasi-BIC. This work demonstrates a double trapezoid structure with strong resonance properties, which has significant implications for exploring the phenomenon of BIC.
		
	\end{abstract}
	
	\maketitle
	
	
	\section{\label{sec:level1}Introduction}
	
Bound state in the continuum (BIC) as an optical phenomenon to limit the loss of light, the phenomenon of which was originated from quantum physics and described as a relationship with a continuous medium threshold. BIC was first proposed by von Neumann and Wigner in 1929 [1], but it’s not demonstrated in quantum systems. It was not until 1992 that Capasso’s group proposed a BIC experiment in quantum systems [2]. Then, many specific structures were used to support BIC, caused a leaky mode with a resonance state of zero width [3]. Until now, there has been a great deal of researches on resonant states, which lie in the continuum with a finite lifetime, i.e., they are born from continuum states and decay eventually. Regulating a structure with specific parameters, resonance states interfered by some different channels yield zero-width resonance, corresponding to BIC without energy loss.
	
Recently, researches proposed symmetrical structures to discuss BICs in photonic crystals or metasurfaces. Grating as a high performance structure have been researched in many fields, and it is suitable as a highly symmetrical structure for the study of BIC. Doeleman et al. stdudied grating’s polarization vortex of BIC in 2018 [4]. In the same year, Azzam et al. proposed a metal grating and discussed hybrid plasmonic-photonic systems [5]. In 2019, Wu et al. discussed BIC by tuning the distance of gap in grating [6]. Lots of applications with the optical properties of BIC, such as optical modulators [7], optical vortices [8-10], nonlinear harmonic generations[11-13] and nonlinear optical devices[14], sensors[15-19] have been designed. For these previous work about gratings, they proposed some simple structures, or they only focused on spectrum of structures without energy bands and topological charges, and they rarely discuss the energy distribution in the momentum space.
	
In this work, we propose a double trapezoid grating as an asymmetric structure to discuss the relationship between BIC and quasi-BIC. For a normal grating, it can store energy in the symmetrical structure without any loss. When we cut off an oblique angle of the grating to achieve symmetry breaking, the energy in the structure will leak into free space. The process corresponds to BIC turning into quasi-BIC, combined with the \textit{Q} factor becomeing finite. A narrow resonance peak appears due to the interference of two channels, and the distribution of electric field will produce a little offset, energies have some loss due to the interference. In the momentum space, topological charges split and couple with the change of grating. Based on guided mode resonance (GMR), the dispersion relations of grating can be theoretically derived, and we can find the position of BIC in energy bands. We focus on changing the inclination angle of the grating, and utilize finite element method (FEM) to calculate reflection spectrum, from which we find a Fano peak. Our structure can produce a large phase change at the position of Fano peak, which can be used to achieve a large Goos-Hänchen (GH) shift. Then we proposed a high sensitivity sensor based on double trapezoidal grating, it’s caused by the change of refractive index of surrounding medium. Comparing to other works about sensors, the sensitivity of double trapezoidal grating in our work can reach to 349.885 $\mu$m/RIU. Our work provides many potential applications, such as optical switches, high-performance sensors and wavelength division multiplexers.
		\begin{figure*}
	\includegraphics[width=.7\textwidth]{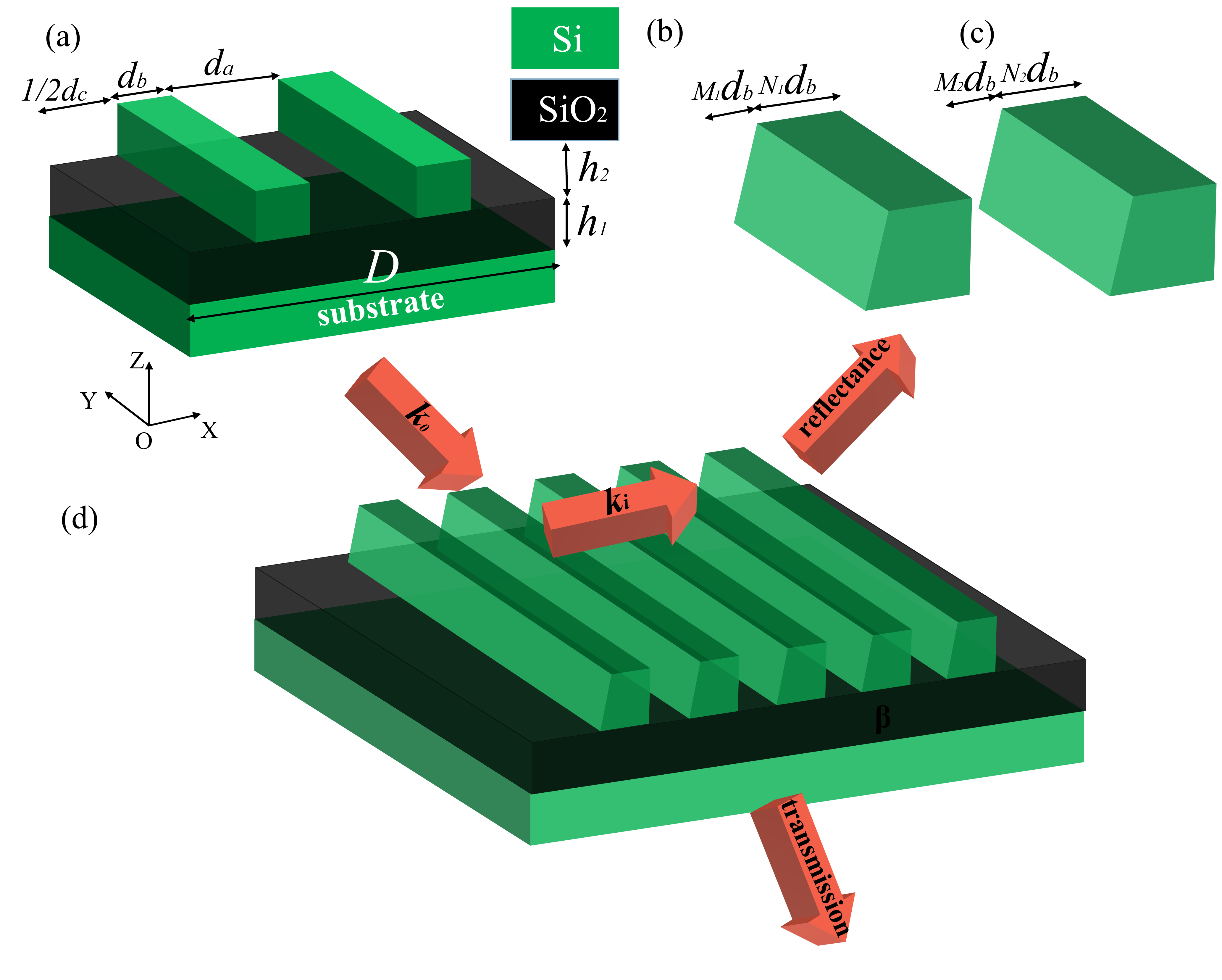}
	\caption{\label{fig:wide} (a) Three-dimensional schematic diagram of the grating structure. (b)-(c) Two trapezoidal grating are controlled by asymmetric geometric parameters $M_{1}$ and $M_{2}$. (d) Schematic diagram of the wave propagation structure satisfying the GMR condition, where $k_{i}$ and $\beta$ are the tangential component of the wave vector.}
\end{figure*}
	\section{BIC in the trapezoid grating and the analysis of dispersion relations}
	We firstly discuss rectangular grating in a periodic photonic crystal slab as shown in Fig. 1(a).The structure maintains a high symmetry with BIC. Then, we cut off one corner of the rectangular grating to break the symmetry shown in Fig. 1(b), where the quasi-BIC can be found. When another rectangle is also damaged in Fig. 1(c), a narrow Fano Peak can be observed with limitation of light field mode. The unite cell of grating can be divide into three parts, grating layer, waveguide layer and substrate layer. The grating layer is chosen to be \textit{Si} and the waveguide layer is $SiO_{2}$. The refractive index of \textit{Si} and $SiO_{2}$ are 3.48 and 1.46, the period length of the whole unite grating D=1544nm and the thickness of waveguide layer is $h_{1}=2000nm$. For the grating layer, the air gap splits into two parts $d_{c}=d_{a}=0.23(1-x)D$. Here, the rectangular grating has a fixed parameter $d_{b}=0.27D$ and the height of grating layer is $h_{2}=500nm$. Extracting the trapezoidal grating separately as shown in Figs. 1(b) and (c), we define parameters $M_{1}$ and $M_{2}$ to control tilt angle, the rest are controlled by N1 and N2, which satisfy $M_{1}+N_{1}=1 (M_{2}+N_{2}=1)$.

	Based on guided mode resonance (GMR) excitation of the grating waveguide layer, we can give a schematic representation of the dispersion relation. The propagation diagram can be seen in Fig. 1(d). In the system, tangential component of wave vector is defined as ki, angular frequency is $\omega$ and \textit{c} is light speed in the air. For the background of air, we have $k_{i}=k_{0}sin\theta$  with the wave vector in air $k_{\theta}=\omega/c$.
	
	\begin{figure*}
		\includegraphics[width=.9\textwidth]{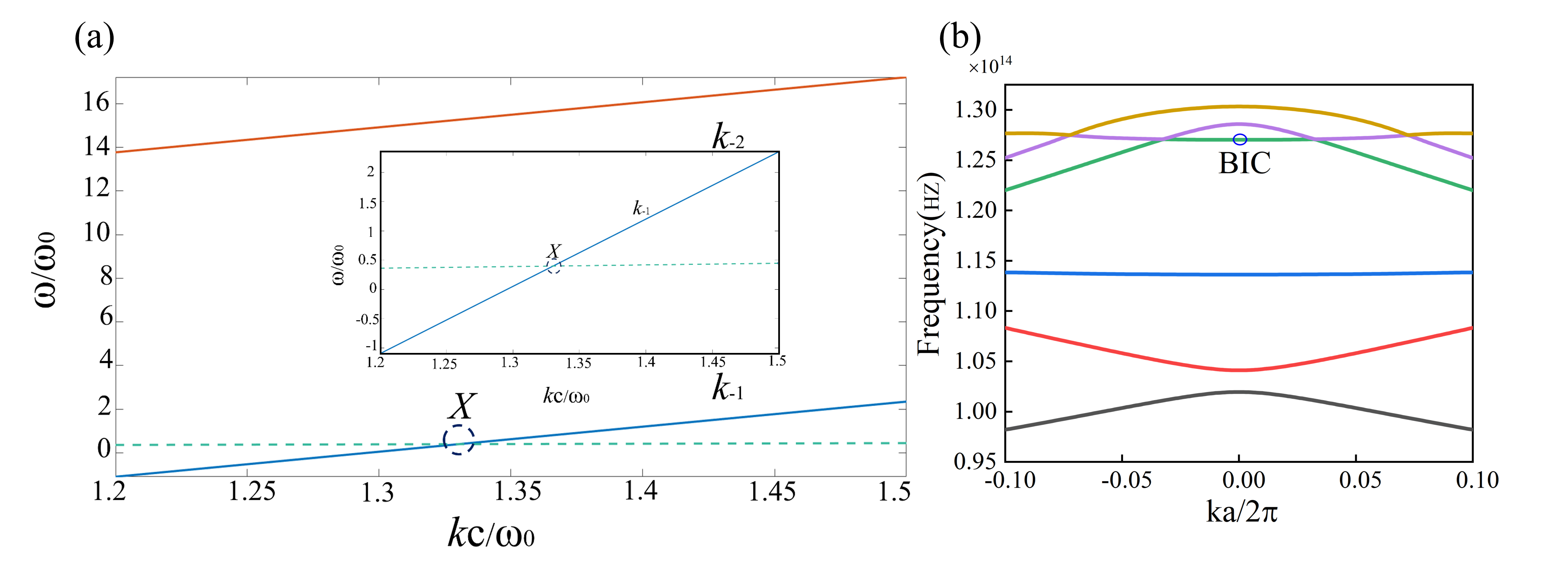}
		\caption{\label{fig:wide} (a) The dispersion relation of waveguide layer represents gray dotted line, the blue solid line and the green realization represent the components of the wave vector in the waveguide layer, where the inset shows an enlarged view of the intersection. (b) In the first Brillouin zone, the range of $ka/2\pi$ is from -0.1 to 0.1, and the point of BIC is found in energy bands.}
	\end{figure*}
	
When the light beam in grating, a portion of the light passes through the grating and the other part is trapped in the waveguide layer. When the resonance effect is generated with leakage disappeared, there is no light transmitted and a sharp resonance peak can be observed. In the grating, we define $k_{i}=k_{0}-i·2\pi/D$, $k_{0}=2\pi/\lambda$, where $i=\pm1, \pm2\dots$, and $2\pi/D$ is the basic vector of the reciprocal lattice. Since propagating constant $\beta$ equals $k_{i}$ in the guide mode, then the expression for GMR can is expressed as
	\begin{eqnarray}
		k_{i}=k_{0}sin\theta-i\frac{2\pi}{D}=\beta
	\end{eqnarray}

Then the difference of the modal propagation coefficient $\beta_{x}$ can be written as [15]
	\begin{eqnarray}
		\Delta\beta_{x}=\frac{2\pi n_{wg}}{\lambda}\Delta(sin\theta)
	\end{eqnarray}
	
where the difference of $\theta$ is given by
	\begin{eqnarray}
\Delta(sin\theta)=\frac{\Delta\lambda}{n_{wg}D}
	\end{eqnarray}
	
	Substituting Eq. (3) into Eq. (2), based on the Lorentzian model with the assumption, the resonance condition is explained as a pole in the complex domain [20], i.e., $\Delta\beta_{x}=\gamma$, where $\gamma$ is the coupling loss coefficient. Then we can derive
	\begin{equation}
\Delta\lambda=\frac{\lambda D\gamma}{2\pi}
	\end{equation}

Finally, the full width at half-maximum (FWHM) is calculated by
	\begin{eqnarray}
\Delta\lambda_{FWHM}=\frac{\lambda D\gamma}{\pi}
	\end{eqnarray}
	
We can initially calculate the half-height width of the resonance peak. For the slab waveguide theory, $\beta$ is calculated based on the eigenvalue equation
	\begin{equation}
\nu_{1}h_{1}=i\pi+tan^{-1}(m_{0}/\nu_{1})+tan^{-1}(m_{2}\nu_{1}), \quad i=0,1,2\dots
	\end{equation}
	
where the abbreviations
	\begin{eqnarray}
m_{2}=\alpha_{2}tanh[tanh^{-1}(m_{3}/\alpha_{2})+\alpha_{2}h_{2}],\\
\nu_{1}=(k_{0}^{2}n_{wg}^{2}-\beta_{0}^{2})^{1/2},\\
\alpha_{2}=(\beta_{0}^{2}-k_{0}^{2}n_{eff}^{2})^{1/2},\\
n_{eff}=[n_{g}^{2}f+1-f]^{1/2},\\
m_{0}=(\beta_{0}^{2}-k_{0}^{2}n_{s}^{2})^{1/2},\\
m_{3}=(\beta_{0}^{2}-k_{0}^{2})^{1/2}
	\end{eqnarray}
	
where \textit{i} is the order of waveguide mode. We can determine the location of the resonance peak based on the above equations. Considering the incident angle $\theta=7.1^{\circ}$, $\omega_{0}=2\pi c/h_{1}$ and $TE_{0}$ guided mode, the dispersion relation in the waveguide layer can be expressed as
	\begin{multline}
	h_{1}\sqrt{k_{0}^{2}n_{wg}^{2}-\beta^{2}}=atan(\frac{\sqrt{\beta^{2}-k_{0}^{2}n_{0}^{2}}}{k_{0}^{2}n_{wg}^{2}-\beta^{2}})\\
	+atan(\frac{\sqrt{\beta^{2}-k_{0}^{2}n_{s}^{2}}}{\sqrt{k_{0}^{2}n_{wg}^{2}-\beta^{2}}})
	\end{multline}
	
The dispersion relation of waveguide layer corresponds to the gray dotted line in Fig. 2(a). The first and second order tangential wave vector components are blue and green solid lines. The intersection points in the figure we mark as \textit{X}, which means GMR condition caused by $k_{i}$ and $\beta$. Besides, we give the energy bands of grating to match the GMR condition and Fano peak in the next section.
	
Figure 2(b) shows the energy bands of structure. For the periodic structure, we set the size of first Brillouin zone to be $2\pi/D$ and find the point of BIC. In the next work, we calculate reflectance and  transmission spectra with finite element method (FEM) [14]. After the model analysis, we can obtain the enhanced electric field modes with the different tilt angle. When the grating maintains rectangle, light is limited due to limitation of the bound continuous. Then we break the symmetric rectangular grating, a little diffusion of the original light field occurrs. In the following, we will further discuss the change of the models and topological polarization by COMSOL Multiphysics.

	\section{Simulation results}
	In the last few years of reported works, discussing the symmetry of the structure is a way to study BIC. For grating as an optical device, it is favorable to study the BIC phenomenon. In our work, firstly, we consider a normal grating structure, which corresponds to the BIC state. However, some optical phenomena cannot be observed due to the infinite \textit{Q}-factor. After that, we try to cut off an oblique angle of the rectangular grating. When the planar symmetry of the structure is broken, a steep reflection peak is found. We keep one of the rectangle unchanged, and then define an asymmetric parameter $M_{1}$ to control the trapezoidal grating. In order to achieve a better Fano line shape, we need to scan the parameter $M_{1}$ from 0 to 1. We plot the distribution of reflection as a function of the asymmetric parameter $M_{1}$ and wavelength in Fig. 3(a). We have circled the location of the quasi-BIC at the wavelength $\lambda=2387nm$ in the Fig. 3(a) and the narrowest Fano peak occurs when $M_{1}=0.3$. Therefore, we set the tilt of one of the rectangles to 0.3, then control another rectangle.
	
	We adjust the other trapezoid while keeping the parameter $M_{1}$ constant as shown in Fig. 3(b). Similarly, we define a new tilt parameter $M_{2}$, and show the distribution of reflections in Fig. 3(b). We can see the color gradient map, where the red areas represent the location of the peak. In order to get the best Fano peak, we choose $M_{2}=0.5$ in Fig. 3(b). Whether it is greater than 0.5 or less than 0.5, none of Fano peaks can become better than the condition $M_{2}=0.5$. We refer to this situation as quasi-BIC, which is formed by the interference of two channels [21]. The two channels are discrete spectrum narrowband originated from GMR and the other is continuous spectrum broadband provided by resonance in the waveguide layer. In addition, resonance peak will produce a red shift due to that the unit cell varies with $M_{2}$.
	\begin{figure}
		\includegraphics[width=.5\textwidth]{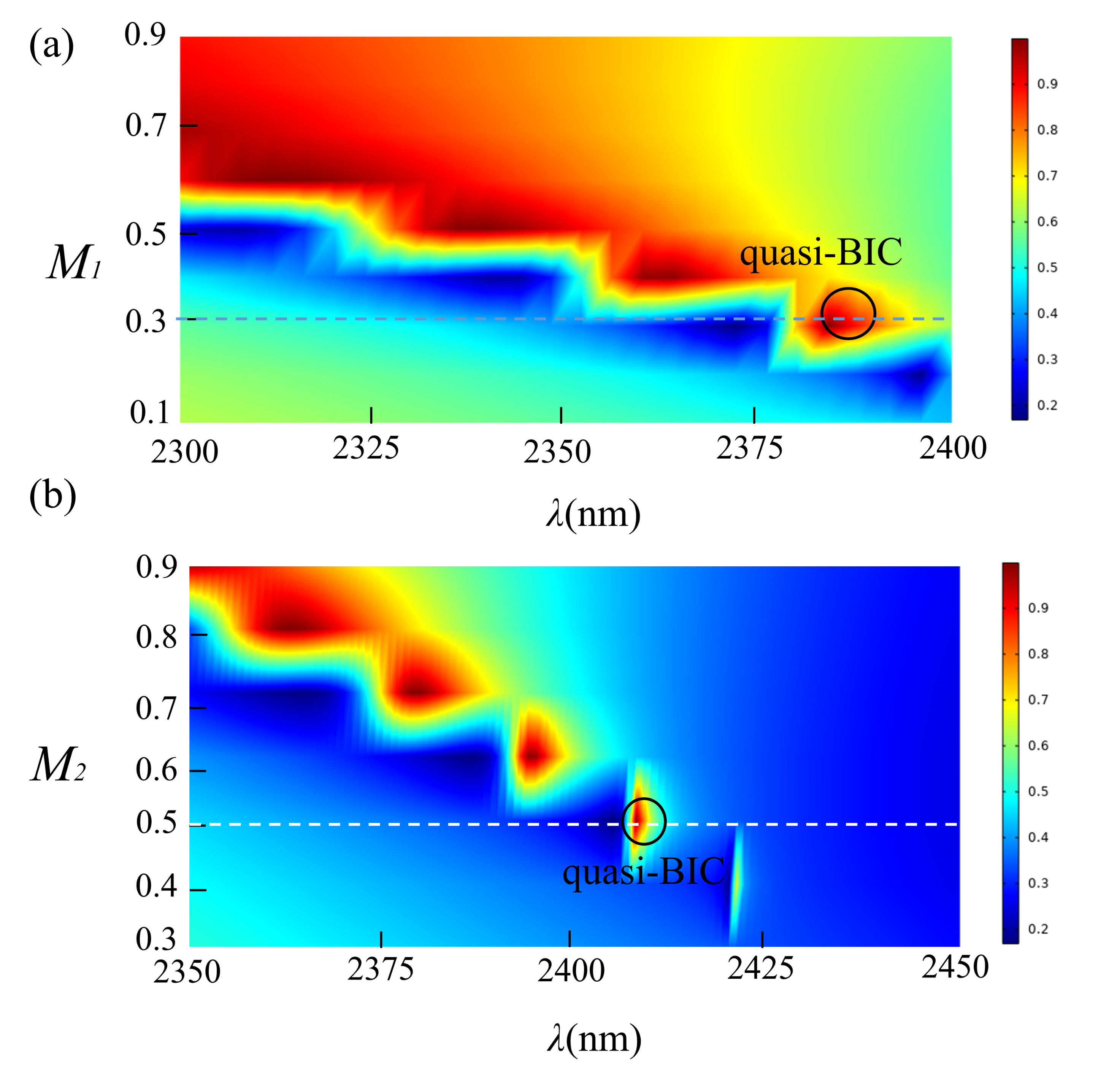}
		\caption{(a) Controlling $M_{1}$ from 0.1 to 0.9 and keeping one of the rectangular grating unchanged, the best Fano peak can appear when $M_{1}$= 0.3. (b) Changing another grating tilt parameter $M_{2}$, a narrowest Fano peak is achieved for $M_{1}=0.3$, $M_{2}=0.5$.}
	\end{figure}

	The topological natures of BIC can be discussed with the change of grating. We show the distributions of the field polarization states in Figs. 4(b) and (d). In the surface momentum space of the grating, total winding of the polarization position remains 2$\pi$. For a symmetrical rectangular grating, topological charges locate at the point $q=1$ as shown in Fig. 4(b), which corresponds to symmetric protection BIC. The sum of topological charges can be calculated as
	
	\begin{equation}
		q=\frac{1}{2\pi}\oint_{L}d_{\psi}=\frac{n_{w}}{2}
	\end{equation}
	
	where the \textit{L} is a closed loop that enclosesg one BIC , $\psi$ is the orientation angle, and $n_{w}$ is defined as the winding number. When the planar symmetry of the structure is broken, the topology integer charge $q=1$ will split into two half topology in Fig. 4(d). Then they will move in the momentum space, with one being left-handed circularly polarized (blue) and the other being right-handed circularly polarized (red). As a result, the field of grating exhibits polarization diversity, where elliptical polarizations, linear and circular polarizations appear near the center of the first Brillouin Zone.

	\begin{figure}
		\centering
		\includegraphics[width=8cm]{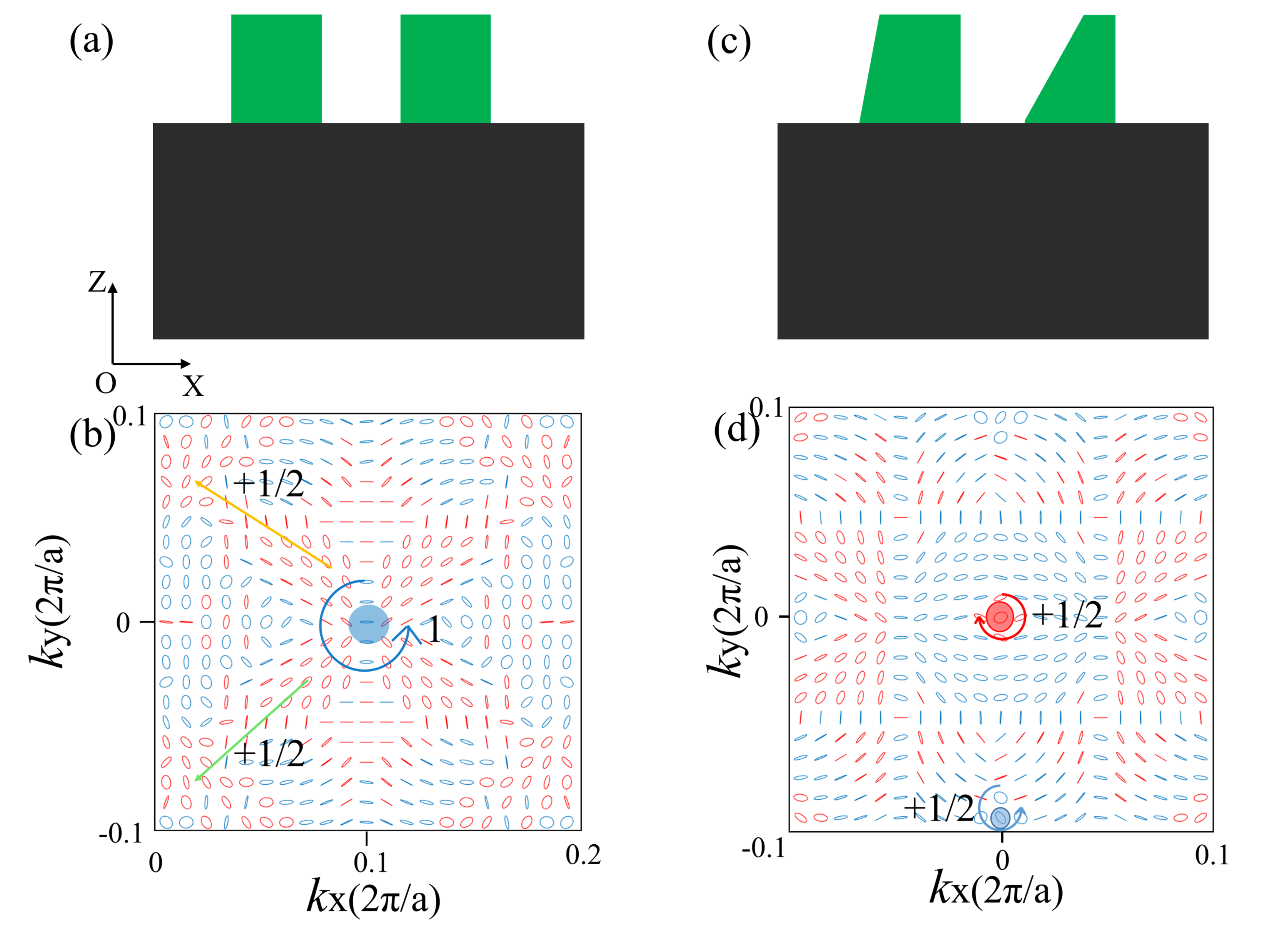}
		\caption{(a) Cross sectional view of a symmetrical rectangular grating. (b) Momentum space polarization with a symmetrical rectangular structure. (c) Cross sectional view of a trapezoidal grating, two trapezoids with different degrees of inclination. (d) Topological charges divided into two half topology with symmetry breaking structures.}
	\end{figure}

		\begin{figure*}
	\includegraphics[width=.9\textwidth]{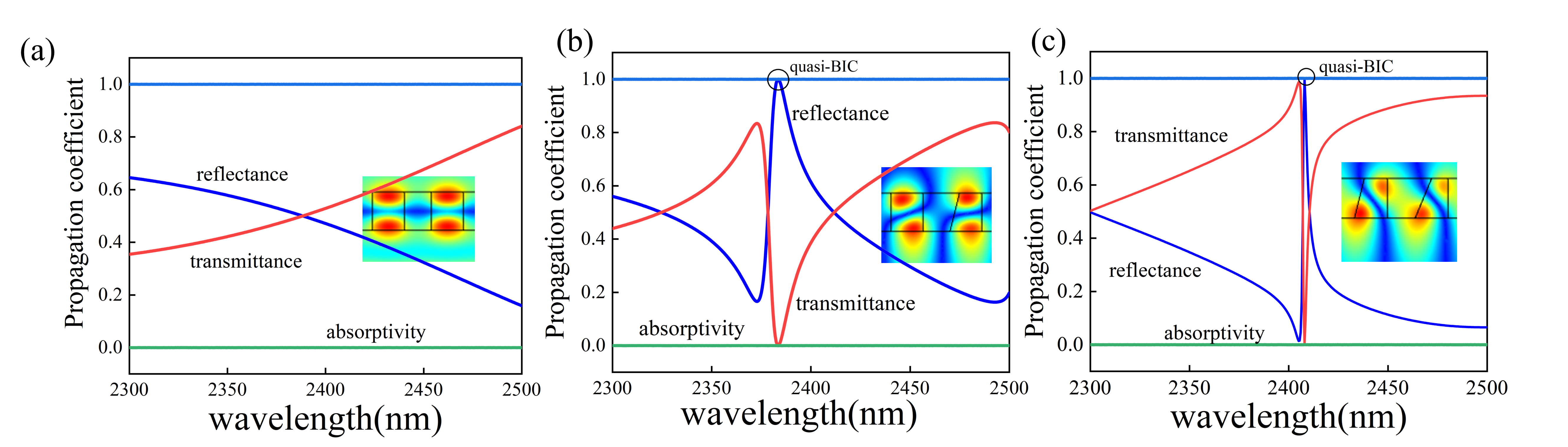}
	\caption{(a) The spectra of symmetrical rectangular grating, reflection is a smooth curve without any peak protrusion, and the electric field pattern is symmetrically distributed. (b) When one of the rectangles is cut into a trapezoid, the original smooth curve appears a steep Fano peak at wavelength equal to 2387nm, a little rotation has occurred in the electric field pattern. (c) Grating is optimized further by control another tilt parameter $M_{2}$, where an ultranarrow Fano peak is found.}
\end{figure*}

	Here, we have calculated reflection and transmission spectra based on finite element method (FEM). For the asymmetric grating structure, the inclination of one of the trapezoids is taken as 0.3, and the other inclination of trapezoids is taken as 0.5. An asymmetrical Fano line can be found in Figs. 5(b) and (c), where their peak reflection efficiency is extremely high and reach to 1, and the intensity distributions of the electric field along the y-axis are also calculated and shown in the insets. Firstly, we discuss the primitive case with $M_{1}=0$ and $M_{2}=0$, which corresponds to the symmetrical protection structure with BIC. Interestingly, when the structure maintains perfect symmetry, the reflection curve is a smooth line without any steep peak, and the resonance width vanishes completely in this case. In the case of BIC, discrete and continuous state will mutually couple in the structure, and its energy will be confined in the grating without any leakage. BIC always means an infinite \textit{Q} factor, and an infinite narrow line width is appeared in the spectrum, which agree with the phenomenon in spectrum of Fig .5(a). The mode of electric field can be seen in the inset of Fig .5(a), where it exhibits a neat concentration of energy.
	
				\begin{figure*}
		\includegraphics[width=.9\textwidth]{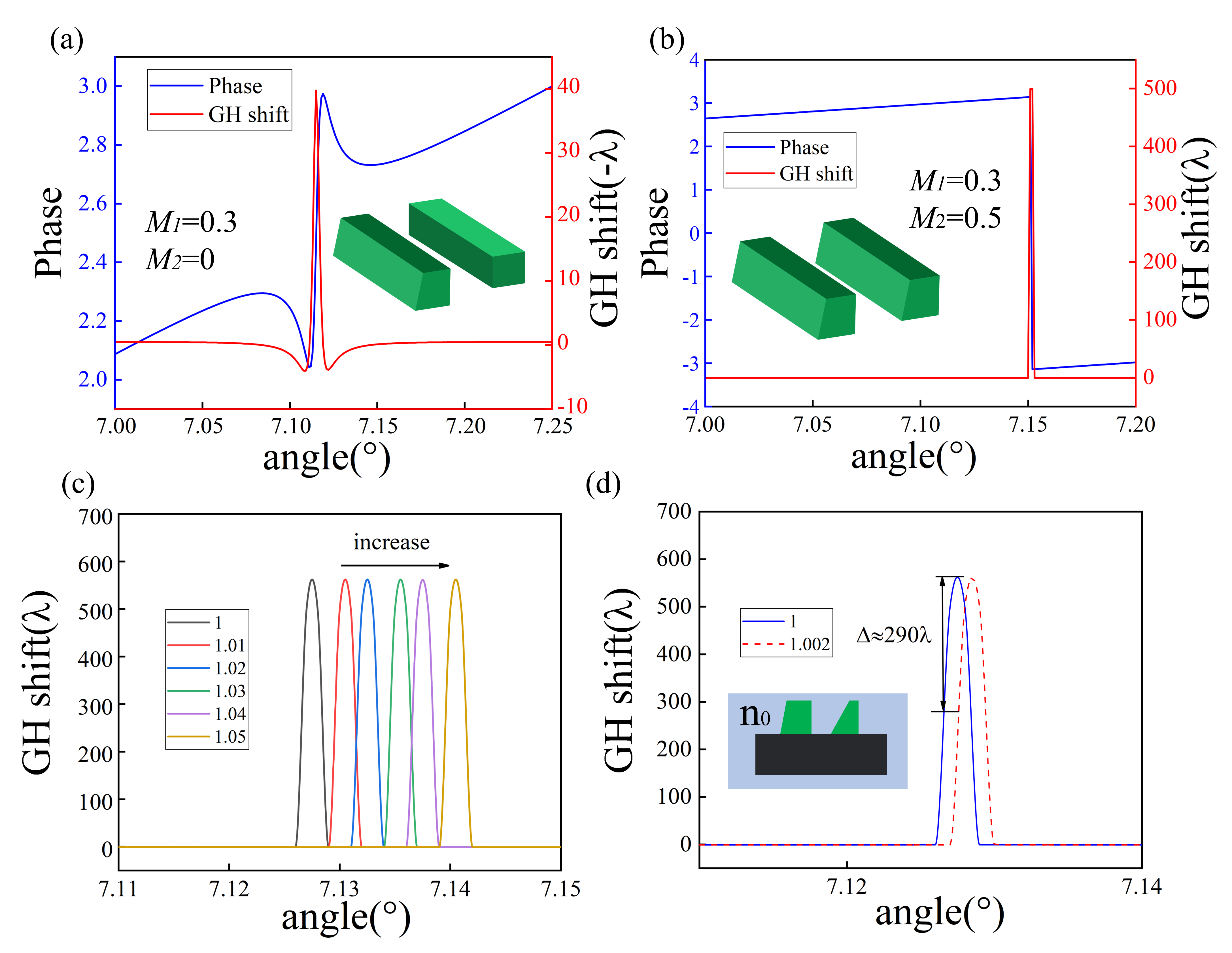}
		\caption{(a) Phase variation with angle for a grating consisting of a trapezoid and a rectangle (blue). GH shift angular spectrum with the asymmetric geometric parameter set to $M_{1} = 0.3$(red), where the incident angle is equals to $7.12^{\circ}$. Phase variation with angle for double trapezoid grating (blue). GH shift angular spectrum with the asymmetric geometric parameter set to $M_{1} = 0.3$ and $M_{2}=0.5(red)$, where incident angle is equals to $7.15^{\circ}$. (c) The dependence of the sensitivity on the refractive index, with the increase of refractive index, the peak of GH shift shifts towards larger angles. (d) GH shift angular spectrum of sensor at $n_{0}=1$ and $n_{0}=1.002$.}
	\end{figure*}

	However, BIC is difficult to be applied and designed in the field of optics due to its infinite \textit{Q} factor and difficulty in manufacturing process. Therefore, considering quasi-BIC is an important way to achieve various good optical properties. Quasi-BIC is a state that is infinitely close to BIC so that it always has an ultra-high \textit{Q} factor and an ultra-narrow Fano peak as can be seen in the spectrum. In the Fig. 5(b), we can find an asymmetrical Fano peak at the wavelength equal to 2387nm, which corresponds to the blue dotted line in Fig. 3(a). Compared to the initial structure, when the one of the rectangles is cut off a corner, the mode of electric fields will produce leakage and offset due to the symmetry broken. Cutting off a corner of the grating leads to a change in structure, topological charges start splitting from $q=1$ in the momentum space.
	
	 Only changing the tilt of one grating in a unit cell, we can get an asymmetrical narrow Fano peak, but it is not the optimal structure to achieve various optical applications. Thus, this structure needs to be optimized further by controlling another inclination angle in a cycle. When the tilt parameters $M_{1}=0.3$ and $M_{2}=0.5$, the Fano line width can be shrunk to a minimum. In Fig. 5(c), an ultra-narrow Fano peak can be reached, the line width is much narrower compared to that in Fig. 5(b). The result of peak is benefit from the optimization of the structure, this line of which corresponds to the white dotted line in Fig. 3(b). The reflection peak obtained is close to complete reflection at wavelength equal to 2410nm. It’s a state that closest to BIC, where topological charges have been splitting into two half charges totally in the momentum space, and it’s electric field energy can be seen in the inset of Fig. 5(c) with a strong energy bondage. In this case, the mode represents leak and loss and the radiation will propagate into free space, which leads to a finite lifetime rather than infinite lifetime of the BIC.
	
	The tilted grating has an ultra-narrow Fano peak, which can achieve a large GH shift. Next, we utilize the quasi-BIC to achieve a large GH shift based on a phase change. The asymmetry geometric parameter $M_{1}$ and $M_{2}$ determine the value of the GH shift, i.e., resonance will become stronger and result in a larger GH shift for specific values of the asymmetric parameter. Here, for the stationary phase method, the lateral GH shift for the reflected and transmitted beams can be calculate by using the expression [22-25],
	\begin{equation}
	S_{GH}=-\frac{\lambda}{2\pi}\frac{\partial_{\varphi_r}}{\partial_\theta}
\end{equation}

\noindent 
where $\varphi_r$ is the reflection phase and the GH shift is proportional to the partial derivative of the reflection phase. Here, we firstly discuss the structure that is only cut off one inclination angle. The phase can change with the incident angle increase under $\lambda$ = 2387 nm and the GH shift angular spectrum is shown in Fig. 6(a). The GH shift shows significantly different behavior around the resonance angle of $\theta = 7.12^{\circ}$. For the double trapezoid structure, the phase change with the incident angle is calculated in Fig. 6(b), which undergoes a phase abrupt at the incident angle $\theta= 7.15^{\circ}$. The GH shift in the Fig. 6(b) is far larger than in the Fig .6(a). In contrast with traditional GH shift enhanced by transmission-type resonances, the maximum GH shift assisted by the quasi-BIC locates at a reflectance peak and it can be detected and utilized more easily.

Based on the large GH shift, we design an ultrasensitive sensor with the quasi-BIC, and define $n_{0}$ as refractive index of surrounding medium. For the surrounding environment of the structure, we firstly calculate GH shift with air $n_{0}$=1 at the room temperature. Then we increase the refractive index of surrounding medium from 1, which can be seen in Fig .6(c). As the index of refraction increases, the peak of GH shift will shift toward a larger incident angle. In addition, the sensitivity of sensor can be defined by the value change of GH shift and the refractive index of surrounding medium, i.e., $S(n_{0})=|d(S_{GH})/dn_{0}|$, and the dependence of the sensitivity on the refractive index can be seen in the Fig. 6(d). The blue solid line represents the refractive index of air $n_{0}=1$, when the medium has changed, the GH shift is a red dotted line in Fig. 6(d). The sensitivity is 349.885 $\mu$m/RIU with the refractive change from 1 to 1.002. It indicates that the sensor we designed has a high sensitivity with the change of medium.

	\section{conclusions}
In conclusion, we have discussed a double trapezoid structure with a strong resonance and split of topological charge. In addition, the point of the BIC in the energy bands can be found by finite element method. We use the GMR and dispersion relations to analyze the interaction between light and matter, which causes asymmetric Fano line shape. We break the symmetry by cutting off a corner of the rectangle, and control the degree of tilt by changing parameters $M_{1}$ and $M_{2}$. With the optimization of parameters, a typical asymmetric Fano line-shape’s width can reduce further, where the parameters are $M_{1}=0.3$ and $M_{2}=0.5$. In the momentum space, topological charge will split into two half charge $q=1/2$ with the destruction of symmetry. Assisted by strong resonance with two different channels, a saltatory phase can be obtained, then the maximum GH shift is achieved at the position of saltatory phase. Our work provides a new perspective to realize BIC, which can support many optical applications such as optical switches, high-performance sensors and wavelength division multiplexers.

	\begin{acknowledgments}
		This work is supported by National Natural Science Foundation of China (1211101294); Intergovernmental Science and Technology Regular Meeting Exchange Project of Ministry of Science and Technology of China (CB02-20); Open Fund of State Key Laboratory of Millimeter Waves (K202238).
	\end{acknowledgments}

	\nocite{*}
	
	\bibliography{apssamp}
	
\end{document}